\documentstyle[12pt]{article}

\font\tenrm=cmr10
 at 10truept

\def\al{\alpha}

\def\ga{\gamma}
\def\de{\delta}

\def\th{\theta}

\def\ka{\kappa}

\def\si{\sigma}

\def\ps{\psi}

\def\De{\Delta}

\def\fr#1#2{{{#1} \over {#2}}}

\def\ket#1{|{#1}\rangle}

\def\half{{\textstyle{1\over 2}}}
\def\frac#1#2{{\textstyle{{#1}\over {#2}}}}

\def\lsim{\mathrel{\rlap{\lower4pt\hbox{\hskip1pt$\sim$}}
    \raise1pt\hbox{$<$}}}
\def\gsim{\mathrel{\rlap{\lower4pt\hbox{\hskip1pt$\sim$}}
    \raise1pt\hbox{$>$}}}
\def\sqr#1#2{{\vcenter{\vbox{\hrule height.#2pt
         \hbox{\vrule width.#2pt height#1pt \kern#1pt
         \vrule width.#2pt}
         \hrule height.#2pt}}}}

\newcommand{\beq}{\begin{equation}}
\newcommand{\eeq}{\end{equation}}
\newcommand{\bea}{\begin{eqnarray}}
\newcommand{\eea}{\end{eqnarray}}
\newcommand{\rf}[1]{(\ref{#1})}

\renewenvironment{thebibliography}[1]
 { \rm
   \begin{list}{\arabic{enumi}.}
    {\usecounter{enumi} \setlength{\parsep}{0pt}
     \setlength{\itemsep}{3pt} \settowidth{\labelwidth}{#1.}
     \sloppy
    }}{\end{list}}

\begin{document}

\baselineskip=16pt
\begin{flushright}
{COLBY 00-02\\}
{IUHET 419\\}
{January 2000\\}
\end{flushright}
\vglue 0.5 truein 

\begin{flushleft}
{\bf SEARCHING FOR LORENTZ VIOLATION\\
 IN THE GROUND STATE OF HYDROGEN
\footnote{\tenrm
Presented by R.B. at Orbis Scientiae 1999,
Ft.\ Lauderdale, Florida, December 1999}
\\}
\end{flushleft}

\vglue 0.8cm
\begin{flushleft}
{\hskip 1 truein
Robert Bluhm$^a$, V. Alan Kosteleck\'y$^b$, and Neil Russell$^c$\\}
\bigskip
{\hskip 1 truein
$^a$Physics Department\\}
{\hskip 1 truein
Colby College\\}
{\hskip 1 truein
Waterville, ME 04901 USA\\}
\medskip
{\hskip 1 truein
$^b$Physics Department\\}
{\hskip 1 truein
Indiana University\\}
{\hskip 1 truein
Bloomington, IN 47405 USA\\}
\medskip
{\hskip 1 truein
$^c$Physics Department\\}
{\hskip 1 truein
Northern Michigan University\\}
{\hskip 1 truein
Marquette, MI 49855 USA\\}
\end{flushleft}

\vglue 0.8cm

\noindent
{\bf INTRODUCTION}
\vglue 0.4 cm

The hydrogen atom has a rich history as a testing
ground of fundamental physics
where small differences between theory and experiment have led
to major advances
\cite{H}.
With the advent of optical high-resolution spectroscopy
and tunable dye lasers, 
new tests
of quantum electrodynamics in hydrogen
have become possible.
The two-photon 1S-2S transition is especially suitable
for high-precision tests and metrology because of its
small natural linewidth of only $1.3$ Hz.
This transition has been measured in a
cold atomic beam of hydrogen
\cite{hansch}
with a precision of $3.4$ parts in $10^{14}$.
It has also been observed
in trapped hydrogen
\cite{cesar}
with a precision of about one
part in $10^{12}$.
As experimental techniques advance,
the measurement of the line center to one part in $10^3$
becomes plausible with an ultimate
resolution of one part in $10^{18}$,
making new tests of fundamental theory possible.

The recent production of antihydrogen in experiments
\cite{antiH}
ushers in a new era for testing
fundamental physics by allowing direct high-precision
comparisons of hydrogen and antihydrogen
\cite{antiHrevs}.
Since the CPT theorem predicts that all local relativistic
quantum field theories of point particles are invariant
under the combined operations of charge conjugation C,
parity reversal P, and time reversal T
\cite{cpt,cpt98},
comparisons of the 1S-2S transition in
hydrogen and antihydrogen should provide a new
high-precision test of CPT.
Indeed,
two future experiments at CERN
\cite{cern}
are aimed at making high-resolution spectroscopic
comparisons of the 1S-2S transitions in spin-polarized hydrogen
and antihydrogen confined within a magnetic trap.
The comparisons of the 1S-2S transition should
have relative figures of merit comparable to that 
of the neutral meson system,
which places a bound on the mass difference between the
$K_0$ and $\bar K_0$ at less than 2 parts in $10^{18}$
\cite{pdg}.

In this proceedings,
we first review a recent theoretical analysis we made of
CPT and Lorentz tests in hydrogen and antihydrogen,
which was published in
Ref.\ \cite{bkr3}.
This included investigations of on-going experiments in
hydrogen as well as the proposed experiments at CERN
comparing hydrogen and antihydrogen.
We showed that these experiments can provide
tests of both CPT-preserving
and CPT-violating Lorentz symmetry.
In addition to examining comparisons of 1S-2S transitions,
we suggested other possible experimental signatures
that are sensitive to CPT or Lorentz breaking,
including measurements of the Zeeman hyperfine levels
in the ground state of hydrogen.
Some of these measurements are currently being made
and preliminary results are presented
for the first time in Walsworth's talk
\cite{wals}.

\vglue 0.6 cm
\noindent
{\bf THEORETICAL FRAMEWORK}
\vglue 0.4 cm

Our analysis uses a theoretical framework that describes
CPT- and Lorentz-violating effects in an extension of the
SU(3)$\times$SU(2)$\times$U(1) standard model
and quantum electrodynamics (QED)
\cite{ck}.
The framework originates from the idea of spontaneous
CPT and Lorentz breaking in a more fundamental theory
such as string theory
\cite{kp1,ks}.
Within this framework,
possible violations of CPT and Lorentz symmetry are included
which maintain desirable features of quantum field theory,
including gauge invariance,
power-counting renormalizability,
and microcausality.
The model is highly constrained,
and only a small number of terms are possible.
These terms are controlled by parameters that can be bounded
by experiments.
This framework has been used to analyze neutral-meson experiments
\cite{kp1,ckpv,expt,k98},
baryogenesis
\cite{bckp},
photon properties
\cite{ck,photons},
Penning-trap experiments
\cite{penning,bkr12,dehmelt},
atomic clock comparisons
\cite{kl99},
muon experiments
\cite{bkl00},
and experiments in spin-polarized matter
\cite{bk00}.

To investigate experiments in hydrogen and antihydrogen,
it suffices to work in the context of the QED extension.
The modified Dirac equation for a
four-component spinor field $\ps$
describing electrons and positrons of mass $m_e$
and charge $q = -|e|$ in a Coulomb potential $A^\mu$ is
\beq
\left( i \ga^\mu D_\mu - m_e - a_\mu^e \ga^\mu
- b_\mu^e \ga_5 \ga^\mu - \half H_{\mu \nu}^e \si^{\mu \nu}
+ i c_{\mu \nu}^e \ga^\mu D^\nu
+ i d_{\mu \nu}^e \ga_5 \ga^\mu D^\nu \right) \ps = 0
\quad .
\label{dirac}
\eeq
Here,
natural units with $\hbar = c = 1$ are used,
$i D_\mu \equiv i \partial_\mu - q A_\mu$,
and $A^\mu = (|e|/4 \pi r, 0)$.
The two terms involving the effective coupling constants
$a_\mu^e$ and $b_\mu^e$ violate CPT,
while the three terms involving
$H_{\mu \nu}^e$, $c_{\mu \nu}^e$, and $d_{\mu \nu}^e$
preserve CPT.
All five of these terms break Lorentz invariance.
Since no CPT or Lorentz violation has been observed,
these parameters are assumed to be small.
Free protons are also described by a modified Dirac
equation involving the corresponding parameters
$a_\mu^p$, $b_\mu^p$, $H_{\mu \nu}^p$, $c_{\mu \nu}^p$,
and $d_{\mu \nu}^p$.

A perturbative treatment in the context of relativistic
quantum mechanics is used to examine the bound states
of hydrogen and antihydrogen.
In this approach,
the unperturbed
hamiltonian $\hat H_0$ and its energy eigenfunctions are
the same for hydrogen and antihydrogen.
All of the perturbations in free hydrogen described by
conventional quantum electrodynamics
are identical for both systems.
However,
the interaction hamiltonians for hydrogen and antihydrogen
including the effects of possible
CPT- and Lorentz-breaking are not the same.
These are obtained in several steps
\cite{bkr12},
involving charge conjugation to obtain the
Dirac equation for antihydrogen,
a field redefinition to eliminate additional time
derivatives in the Dirac equation,
and the use of standard relativistic two-fermion techniques
\cite{bs}.

\vglue 0.6 cm
\noindent
{\bf EXPERIMENTS WITH FREE HYDROGEN}
\vglue 0.4 cm

We first consider free hydrogen and antihydrogen in the absence
of external trapping potentials.
Using a description in terms of
the basis states $\ket{m_J,m_I}$,
with $J=1/2$ and $I=1/2$
describing the uncoupled atomic and
nuclear angular momenta,
the leading-order energy corrections can be computed.
The energy shifts at the 1S and 2S levels are
found to be the same.
For hydrogen they are given by
\bea
\De E^{H} (m_J = \pm \half, m_I = \pm \half) =
(a_0^e + a_0^p - c_{00}^e m_e - c_{00}^p m_p)
\quad\quad\quad\quad
\quad\quad\quad
\nonumber \\
\quad\quad\quad
+ \fr {m_J} {|m_J|} (-b_3^e + d_{30}^e m_e + H_{12}^e)
+ \fr {m_I} {|m_I|} (-b_3^p + d_{30}^p m_p + H_{12}^p)
\quad ,
\label{EHJI}
\eea
where $m_e$ and $m_p$ are the electron and proton masses,
respectively.
The corresponding energy corrections for the 1S and 2S states of
antihydrogen $\De E^{ \bar H}$
are obtained from these by letting
$a_\mu \rightarrow - a_\mu$,
$d_{\mu \nu} \rightarrow - d_{\mu \nu}$,
and $H_{\mu \nu} \rightarrow - H_{\mu \nu}$
for both the electron-positron and proton-antiproton coefficients.

The hyperfine interaction couples the electron and proton
or positron and antiproton spins.
The appropriate basis states are then $\ket{F,m_F}$
which are linear combinations of the states $\ket{m_J,m_I}$.
The selection rules for the two-photon 1S-2S transition
are $\De F = 0$ and $\De m_F = 0$.
These selection rules require that the 1S-2S transitions
in free hydrogen and antihydrogen occur
between states of the same spin configurations.
As a result,
the leading-order energy shifts are equal,
and there are no observable leading-order shifts in frequency
in either hydrogen or antihydrogen.

There are,
however,
subleading-order shifts in the 1S-2S frequencies.
These are due to small relativistic corrections of order $\al^2$
times the CPT- or Lorentz-breaking parameters
which are different at the 1S and 2S levels.
For example,
the term proportional to $b_3^e$ results in a frequency shift in the
$m_F = 1 \rightarrow m_{F^\prime} = 1$ transition
relative to that of the $m_F = 0 \rightarrow m_{F^\prime} = 0$ line
(which remains unshifted)
equal to $\de \nu^H_{1S-2S} \approx - \al^2 b_3^e / 8 \pi$.
However,
electron bounds obtained in $g-2$ experiments
\cite{dehmelt}
suggest that $b_3^e$ is sufficiently small so that
$\de \nu^H_{1S-2S}$ would be below the expected
1S-2S line resolution.

\vglue 0.6 cm
\noindent
{\bf EXPERIMENTS WITH TRAPPED HYDROGEN}
\vglue 0.4 cm

The experiments to be performed at CERN will use trapped
hydrogen and antihydrogen in a magnetic field $B$.
We use the conventional labels $\ket{a}_n$, $\ket{b}_n$,
$\ket{c}_n$, and $\ket{d}_n$ in order of increasing energy
to denote the four S-state hyperfine levels of hydrogen
with principal quantum number $n$.
The $\ket{b}_n$ and $\ket{d}_n$ states have proton and
electron spins that are aligned,
while the remaining two states have mixed spin
configurations given by
\beq
\ket{c}_n = \sin \th_n \ket{-\half,\half} +
\cos \th_n \ket{\half,-\half}
\quad ,
\label{c}
\eeq
\beq
\ket{a}_n = \cos \th_n \ket{-\half,\half} -
\sin \th_n \ket{\half,-\half}
\quad .
\label{a}
\eeq
The mixing angles depend on $n$ and obey
$\tan 2 \th_n \approx (51 {\,\, \rm mT})/n^3 B$.

The states $\ket{c}_n$ and $\ket{d}_n$ are
low-field seeking states that remain confined in the trap.
However,
collisional effects lead to a loss of population
over time of the $\ket{c}_n$ states.
One possible measurement would therefore be to compare
the frequencies $\nu_d^H$ and $\nu_d^{\bar H}$ for transitions
between $\ket{d}_n$ states at the 1S and 2S levels.
These measurements are particularly attractive
because the 1S-2S $\ket{d}_1 \rightarrow \ket{d}_2$
transitions are field-independent for small values of $B$.
However,
since the spin configurations of the
1S $\ket{d}_1$ and 2S $\ket{d}_2$ states are the same,
we find no observable frequency shifts to
leading order in this case,
i.e.,
$\de \nu_d^H = \de \nu_d^{\bar H} \simeq 0$.

An alternative experiment would look at transitions
involving the mixed states $\ket{c}_n$ and $\ket{a}_n$.
Here,
the $n$ dependence in the hyperfine splitting
leads to a difference in the amount of
spin mixing at the 1S and 2S levels.
This gives rise to a nonzero frequency shift in 1S-2S
transitions between $\ket{c}_n$ hyperfine states:
\beq
\de \nu_c^H \simeq
-(\cos 2 \th_2 - \cos 2 \th_1)
(b_3^e - b_3^p - d_{30}^e m_e + d_{30}^p m_p - H_{12}^e + H_{12}^p)
\quad ,
\label{nucH}
\eeq
The corresponding transition for antihydrogen can be computed as well.
The hyperfine states in antihydrogen in the same
magnetic fields have opposite spin assignments for the
positron and antiproton compared to those of the electron
and proton in hydrogen.
The resulting shift $\de \nu_c^{\bar H}$ for antihydrogen
is the same as for hydrogen except that the signs of $b_3^e$
and $b_3^p$ are changed.

Two possible experimental signatures for CPT and Lorentz breaking
follow from these results.
The first involves looking for sidereal time variations
in the frequencies $\nu_c^H$ and $\nu_c^{\bar H}$.
The second involves measuring the instantaneous 1S-2S frequency
difference in hydrogen and antihydrogen in the same
magnetic trapping fields.
In either case,
the strength of the signal would depend on the difference
in the amount of spin mixing at the 1S and 2S levels.
The optimal experiment would be one that maximizes the
1S-2S spin-mixing difference,
which is controlled by the magnetic field $B$.
Since the 1S-2S $\ket{c}_1 \rightarrow \ket{c}_2$
transition in hydrogen and antihydrogen is field dependent,
these experiments would need to overcome line broadening
effects due to field inhomogeneities in the trap.

\vglue 0.6 cm
\noindent
{\bf EXPERIMENTS ON THE GROUND-STATE HYPERFINE LEVELS}
\vglue 0.4 cm

The best tests of CPT and Lorentz symmetry in atomic
systems are those that have the sharpest frequency resolutions.
It is therefore natural to consider other transitions
in hydrogen and antihydrogen besides the 1S-2S transition
that can be measured with high precision.
One candidate set involves measurements of
the ground-state hyperfine levels in
hydrogen and antihydrogen.
For example,
hydrogen maser transitions between $F = 0$ and $F^\prime = 1$
hyperfine states can be measured with accuracies of
less than $1$ mHz.
High-resolution radio-frequency measurements can also
be made on transitions between Zeeman hyperfine
levels in a magnetic field.

To examine these types of experiments,
we compute the energy shifts of the four hydrogen
ground-state hyperfine levels in a magnetic field.
The spin-dependent contributions to the energy are
\beq
\De E_a^H \simeq \hat \ka
(b^e_3 - b_3^p - d^e_{30}m_e + d^p_{30}m_p - H^e_{12} + H^p_{12})
\quad ,
\label{ea}
\eeq
\beq
\De E_b^H \simeq
b^e_3 + b_3^p - d^e_{30}m_e - d^p_{30}m_p - H^e_{12} - H^p_{12}
\quad ,
\label{eb}
\eeq
\beq
\De E_c^H \simeq - \De E_a^H
\quad ,
\label{ec}
\eeq
\beq
\De E_d^H \simeq - \De E_b^H
\quad ,
\label{ed}
\eeq
where $\hat \ka \equiv \cos 2\th_1$.

In a very weak or zero magnetic field $\hat \ka \simeq 0$
and the energies of the states $\ket{a}_1$ and $\ket{c}_1$
are unshifted while the states $\ket{b}_1$ and $\ket{d}_1$
acquire equal and opposite shifts.
The degeneracy of the three $F=1$ levels is therefore lifted.
A conventional hydrogen maser operates on the field-independent
transition $\ket{c}_1 \rightarrow \ket{a}_1$ in the presence
of a small ($B \lsim 10^{-6}$ T) magnetic field.
Since $\hat \ka \lsim 10^{-4}$ in this case,
the leading-order effects due to CPT and Lorentz
violation are suppressed.
However,
the frequencies of the Zeeman hyperfine transitions
between $F=1$ levels are affected by CPT and Lorentz
violation and have unsuppresed corrections.
For example,
the correction to the $\ket{c}_1 \rightarrow \ket{d}_1$
transition frequency in a very weak field is given by
\beq
\de \nu_{c \rightarrow d}^{\rm H \, maser} \simeq
(- b^e_3 - b_3^p + d^e_{30}m_e + d^p_{30}m_p + H^e_{12} + H^p_{12})/2\pi
\quad .
\label{delnu}
\eeq
A signature of CPT and Lorentz violation would thus be
sidereal time variations in the frequency
$\nu_{c \rightarrow d}^{\rm H \, maser}$.

The transition $\ket{c}_1 \rightarrow \ket{d}_1$ in a
hydrogen maser is field-dependent,
and one would expect field broadening to
limit the resolution of frequency measurements.
However,
as described by Walsworth
\cite{wals},
it is possible to perform a double-resonance experiment
\cite{andre}
in which variations of
the $\ket{c}_1 \rightarrow \ket{d}_1$ transition
are determined by monitoring their effect on the usual
$\ket{a}_1 \rightarrow \ket{c}_1$ maser line.
This then permits a search for sidereal variations in the frequency
$\nu_{c \rightarrow d}^{\rm H \, maser}$.
Walsworth's group at the Harvard-Smithsonian Center
has begun this experiment,
and their preliminary results indicate that the
sidereal variations in $\nu_{c \rightarrow d}^{\rm H \, maser}$
can be bounded at a level of approximately 0.7 mHz.
This corresponds to a bound on the combination of
parameters in $\de \nu_{c \rightarrow d}^{\rm H \, maser}$
in Eq.\ \rf{delnu}
at a level of $10^{-27}$ GeV.
Defining a figure of merit as the ratio
of the amplitude of the sidereal variations of the energy
relative to the energy itself,
i.e., $r_{\rm hf}^H \equiv (\De E_{\rm hf})_{\rm sidereal}/E_{\rm hf}$,
one obtains from the results of Walsworth's experiment the value
\beq
r_{\rm hf}^H \lsim 10^{-27}
\quad .
\label{rhf}
\eeq
This now gives one of the sharpest bounds
on CPT and Lorentz violation for protons and electrons.

In principle,
measurements of this kind can also be made on
the Zeeman hyperfine levels in antihydrogen.
Since only in a direct comparison of matter and antimatter can
the CPT-violating effects be isolated,
it is hoped that the technical obstacles of performing
radio-frequency spectroscopy in trapped antihydrogen
can be overcome.
As an alternative to measurements in a very weak magnetic field,
which might be hard to maintain in a trapping environment,
one could perform a comparison of $\ket{c}_1 \rightarrow \ket{d}_1$
transitions in hydrogen and antihydrogen at the field-independent
transition point $B \simeq 0.65$ T.
At this field strength,
the electron and proton spins in the $\ket{c}_1$ state
are highly polarized with $m_J = \half$ and $m_I = -\half$.
The transition $\ket{c}_1 \rightarrow \ket{d}_1$ is effectively
a proton spin-flip transition.
The instantaneous difference in this transition for hydrogen
and antihydrogen is found to be
$\De \nu_{c \rightarrow d} \simeq -2 b^p_3/\pi$.
A measurement of this difference would provide a
direct, clean, and accurate test of CPT for the proton.

\vglue 0.6 cm
\noindent
{\bf CONCLUSIONS}
\vglue 0.4 cm

In summary,
we find that by using a general framework we are able
to analyze proposed tests of CPT in hydrogen and antihydrogen.
We find that in addition to testing CPT,
these experiments will also test Lorentz symmetry.
Our analysis shows that in comparisons of 1S-2S
transitions in hydrogen and antihydrogen,
control of the spin mixing at the 1S and 2S levels
is an essential feature in designing an effective test
of CPT and Lorentz symmetry.
We also find that high-resolution radio frequency
experiments in hydrogen or antihydrogen offer
the possibility of new and precise tests of CPT
and Lorentz symmetry.
One very recent experiment using a double-resonance
technique in a hydrogen maser has obtained a
new CPT and Lorentz bound at the level of
$10^{-27}$ for electrons and protons.

\vglue 0.6 cm
\noindent
{\bf ACKNOWLEDGMENTS}
\vglue 0.4 cm

This work was supported in part
by the National Science Foundation
under grant number PHY-9801869.

\vglue 0.6 cm
\noindent
{\bf REFERENCES}
\vglue 0.4 cm


\begin{thebibliography}{xx}

\bibitem{H}
See, for example,
G.F. Bassani, M. Inguscio, and T.W. H\"ansch, eds.,
{\it The Hydrogen Atom}
(Springer-Verlag, Berlin, 1989).

\bibitem{hansch}
T. Udem et al.,
Phys. Rev. Lett. {\bf 79} (1997) 2646.

\bibitem{cesar}
C.L. Cesar et al.,
Phys. Rev. Lett. {\bf 77} (1996) 255.

\bibitem{antiH}
G. Baur et al.,
Phys. Lett. B {\bf 368} (1996) 251;
G. Blanford et al.,
Phys. Rev. Lett. {\bf 80} (1998) 3037.

\bibitem{antiHrevs}
M. Charlton, J. Eades, D. Horvath, and C. Zimmermann,
Phys. Reps. {\bf 241} (1994) 65;
J. Eades, ed.,
{\it Antihydrogen}
(J.C. Baltzer, Geneva, 1993).

\bibitem{cpt}
See, for example,
R. G. Sachs,
{\it The Physics of Time Reversal},
University of Chicago, Chicago, 1987.

\bibitem{cpt98}
V.A. Kosteleck\'y, ed.,
{\it CPT and Lorentz Symmetry}
(World Scientific, Singapore, 1999).

\bibitem{cern}
B. Brown et al.,
Nucl. Phys. B (Proc. Suppl.) {\bf 56A} (1997) 326;
M.H. Holzscheiter et al.,
Nucl. Phys. B (Proc. Suppl.) {\bf 56A} (1997) 336.

\bibitem{pdg}
See, for example,
R.M. Barnett {\it et al.},
Review of Particle Properties,
Phys. Rev. D {\bf 54} (1996) 1.

\bibitem{bkr3}
R. Bluhm, V.A. Kosteleck\'y and N. Russell,
Phys. Rev. Lett. {\bf 82} (1999) 2254.

\bibitem{wals}
R. Walsworth {\it et al},
in preparation;
see also this volume.

\bibitem{ck}
D. Colladay and V.A. Kosteleck\'y,
Phys. Rev. D {\bf 55} (1997) 6760;
Phys. Rev. D {\bf 58}, 116002 (1998).

\bibitem{kp1}
V.A. Kosteleck\'y and R. Potting,
Nucl. Phys. B {\bf 359} (1991) 545;
Phys. Lett. B {\bf 381} (1996) 389.

\bibitem{ks}
V.A. Kosteleck\'y and S. Samuel,
Phys. Rev. Lett. {\bf 63} (1989) 224;
ibid. {\bf 66} (1991) 1811;
Phys. Rev. D {\bf 39} (1989) 683;
ibid. {\bf 40} (1989) 1886.

\bibitem{ckpv}
V.A. Kosteleck\'y and R. Potting,
in D.B. Cline, ed.,
{\it Gamma Ray--Neutrino Cosmology and Planck Scale Physics} \rm
(World Scientific, Singapore, 1993)
(hep-th/9211116);
Phys. Rev. D {\bf 51} (1995) 3923;
D. Colladay and V. A. Kosteleck\'y,
Phys. Lett. B {\bf 344} (1995) 259;
Phys. Rev. D {\bf 52} (1995) 6224;
V.A. Kosteleck\'y and R. Van Kooten,
Phys. Rev. D {\bf 54} (1996) 5585.

\bibitem{expt}
B. Schwingenheuer {\it et al}.,
Phys. Rev. Lett. {\bf 74} (1995) 4376;
L.K. Gibbons et al.,
Phys. Rev. D {\bf 55} (1997) 6625;
KTeV Collaboration,
presented by Y.B. Hsiung at the KAON 99 conference,
Chicago, June 1999;
OPAL Collaboration, R. Ackerstaff
{\it et al.},
Z. Phys. C {\bf 76} (1997) 401;
DELPHI Collaboration,
M. Feindt {\it et al.},
preprint DELPHI 97-98 CONF 80 (July 1997).

\bibitem{k98}
V.A. Kosteleck\'y,
Phys. Rev. Lett. {\bf 80} (1998) 1818;
Phys. Rev. D {\bf 61}, 016002 (2000).

\bibitem{bckp}
O. Bertolami et al.,
Phys. Lett. B {\bf 395}, 178 (1997).

\bibitem{photons}
S.M. Carroll, G.B. Field, and R. Jackiw,
Phys. Rev. D {\bf 41} (1990) 1231;
R. Jackiw and V.A. Kosteleck\'y,
Phys. Rev. Lett. {\bf 82} (1999) 3572;
M. P\'erez-Victoria,
Phys. Rev. Lett. {\bf 83} (1999) 2518;
J.-M. Chung,
Phys. Lett. B {\bf 461} (1999) 138.

\bibitem{penning}
P.B. Schwinberg, R.S. Van Dyck, Jr., and H.G. Dehmelt,
Phys. Lett. A {\bf 81} (1981) 119;
R.S. Van Dyck, Jr., P.B. Schwinberg, and H.G. Dehmelt,
Phys. Rev. D {\bf 34}
(1986) 722; L.S. Brown and G. Gabrielse,
Rev. Mod. Phys. {\bf 58} (1986) 233;
R.S. Van Dyck, Jr., P.B. Schwinberg, and H.G. Dehmelt,
Phys. Rev. Lett. {\bf 59} (1987) 26;
G. Gabrielse et al.,
Phys. Rev. Lett. {\bf 74} (1995) 3544;

\bibitem{bkr12}
R. Bluhm, V.A. Kosteleck\'y and N. Russell,
Phys. Rev. Lett. {\bf 79} (1997) 1432;
Phys. Rev. D {\bf 57} (1998) 3932.

\bibitem{dehmelt}
R.K. Mittleman, I.I. Ioannou, H.G. Dehmelt, and N. Russell,
Phys. Rev. Lett. {\bf 83} (1999) 2116;
H.G. Dehmelt, R.K. Mittleman, R.S. Van Dyck, Jr., and P. Schwinberg,
Phys. Rev. Lett. {\bf 83} (1999) 4694.

\bibitem{kl99}
V.A. Kosteleck\'y and C.D. Lane,
Phys. Rev. D {\bf 60}, 116010 (1999).

\bibitem{bkl00}
R. Bluhm, V.A. Kosteleck\'y and C.D. Lane,
Phys. Rev. Lett. {\bf 84} (2000) 1098.

\bibitem{bk00}
R. Bluhm and V.A. Kosteleck\'y,
Phys. Rev. Lett. {\bf 84} (2000) 1381.

\bibitem{bs}
See, for example,
H.A. Bethe and E.E. Salpeter,
{\it Quantum Mechanics of One- and Two-Electron Atoms}
(Plenum, New York, 1957).

\bibitem{andre}
H.G. Andresen,
Z. Phys. {\bf 210} (1968) 113.


\end{thebibliography}
\end{document}